\documentclass[prl,twocolumn,showpacs,amsmath,amssymb,superscriptaddress]{revtex4}
\usepackage{graphicx}
\usepackage{bm}
\begin{document}

\title{Quantum Lifshitz point in the infinite dimensional Hubbard model}

\author{F. G\"{u}nther}
\author{G. Seibold}
\affiliation{Institut f\"{u}r Physik BTU Cottbus, P.O. Box 101344, 03013
  Cottbus}
\author{J. Lorenzana}
\affiliation{SMC-INFM, ISC-CNR, Dipartimento di Fisica,
Universit\`a di Roma La Sapienza, P. Aldo Moro 2, 00185 Roma, Italy}
\date{\today}
\begin{abstract}
We show that the 
Gutzwiller variational wave function is surprisingly accurate for the
computation of magnetic phase boundaries in the infinite dimensional
Hubbard model. This allows us to substantially extend known 
phase diagrams.  For both the half-hypercubic and the hypercubic
lattice a large part of the
phase diagram is occupied by an incommensurate phase, intermediate 
between the ferromagnetic and the
paramagnetic phase. In case of the 
hypercubic lattice the three phases join at a new quantum Lifshitz
point at which the order parameter is critical and the stiffness
vanishes.
\end{abstract}

\maketitle

The Hubbard model was originally introduced to study
ferromagnetism in strongly correlated metals \cite{hub63,gut63,kan63}.  
This phenomenon as prototypically realized in Fe, Co
and Ni is one of the oldest phenomena investigated in solid state theory.
A related problem is incommensurate spin-density-wave order as
manifested in Cr and its alloys \cite{faw94}.
Also various transition metal oxides show incommensurate  
magnetic phases which are often accompanied by charge order,
like cuprates \cite{tra95}, 
nickelates \cite{hayden92} and manganites\cite{mor98}.

Despite decades of investigation not much is known about the magnetic
phase diagram of the Hubbard model. The simplest treatments \cite{mah00,faz90}
partition the zero temperature phase diagram in three regions: 
antiferromagnetic (AFM) close to $n=1$ particles per atom,
ferromagnetic (FM) at
large interaction and far from $n=1$ and paramagnetic (PM) at small
interaction and/or close to $n=0,2$. More sophisticated
treatments include in addition incommensurate (IC) 
phases \cite{pen66,don95}.

 Infinite dimensional lattices offer an unique opportunity to study
the competition between PM and FM keeping the
problem tractable and, at the same time, retaining much of the physics
expected in three dimensional lattices \cite{geo96}.
Here we investigate the  Hubbard model in the
limit of infinite dimension ${\cal D}$ using the Gutzwiller 
variational wave function (GWF) \cite{gut65}. Instabilities of the FM and
PM ground state are systematically studied as a
function of momentum, doping, and interaction strength 
using a random-phase-approximation (RPA) like
expansion \cite{sei01,sei03,sei04b}. In principle the
model can be solved exactly in this limit
 by using dynamical mean-field theory (DMFT) which maps the problem 
to an impurity problem amenable of numerical solution \cite{geo96}. 
Limitations on the numerical
algorithm, however,  have prevented for an extensive zero temperature
investigation of the phase diagram. A study by Uhrig on the stability of the 
FM phase in an infinite dimensional generalization 
of the fcc lattice, the so called half-hypercubic (hhc) lattice, is   
one of the few cases where the $T=0$ self-consistent DMFT
problem has been solved exactly \cite{uhr96}.

Here we show, comparing with exact results when available, that the 
celebrated GWF is surprisingly accurate for the determination of 
magnetic phase boundaries in infinite dimensional lattices. 
In addition we significantly extend the computation of the $T=0$  magnetic
phase diagram to regions in parameter space yet poorly explored by 
DMFT methods. More specifically, 
for the hhc, we show that the GWF FM instability line agrees 
with the exact computation by Uhrig \cite{uhr96}. In addition 
we present the first systematic investigation of the stability of the
PM state in this lattice and find that the PM and FM never meet but
are separated by an IC phase (Fig.~\ref{hhc}). 
A similar systematic study of the hypercubic lattice (hc) shows also 
a large region of, yet poorly studied, IC order (Fig.~\ref{hc}). 
In contrast to the hhc case, PM and FM phases have a common phase boundary. 
The IC phase, the FM phase and the PM phase join in a quantum version of the 
multicritical Lifshitz point (LP).  Classical LP are interesting because  they are
associated with unusual critical exponents \cite{hor80}. We
anticipate unusual critical behavior for a quantum LP (QLP) shall it
occur at physical dimensions.

We start from the  Hubbard Hamiltonian \cite{hub63,gut63,kan63} 
\begin{eqnarray}\label{hub01}
&H &=\sum_{i,j,\sigma}
     t_{ij}\hat{c}^\dagger_{i\sigma}\hat{c}_{j\sigma}
     +U\sum_{i}\hat{n}_{i\uparrow}\hat{n}_{i\downarrow}.
\end{eqnarray}
where $\hat{c}^\dagger_{i\sigma}$ ($\hat{c}_{i\sigma}$) destroys (creates) an electron with spin $\sigma$ at site $i$. 
$U$ is the on-site Hubbard interaction and $t_{ij}$ denotes the
hopping parameter between sites $i$ and $j$. 
  
The energy of the model is evaluated within the GWF for polarized and
unpolarized phases.  In the limit ${\cal D}\rightarrow \infty$ the 
Gutzwiller approximation (GA) to the 
variational problem becomes exact\cite{met88} which greatly simplifies 
the computations. The stability of these solutions is studied by
an RPA analysis at different  momenta.   By construction, the 
limits of stability thus obtained correspond to the point where a 
more stable
variational solution would be found in a fully unrestricted computation of
the GWF energy. 
This requires a rotationally invariant formulation of the GA energy
functional \cite{li89} and to take into account the change in the
double occupancy for small quasistatic changes of the charge and spin 
distribution as in Vollhardt's GA-based Fermi liquid
computation \cite{vol84}. Basically we compute the dynamic
spin susceptibility 
$\chi_q(\omega)=-\frac{1}{N} \int e^{i\omega t}
\langle {\cal T} S^+_q(t) S^-_{-q}(0)\rangle$
in both para- and FM states which can be obtained from
a density expansion of the Gutzwiller energy functional. 
Details of the formalism are given in Ref.~\cite{sei04b}. 
For simplicity we neglect macroscopic phase separation \cite{faz90,don95} 
which, in any case, would be frustrated if the 
long-range Coulomb interaction were taken into account \cite{lor02}.
  
\begin{figure}[tbp]
\begin{center}
\includegraphics[width=8cm,clip=true]{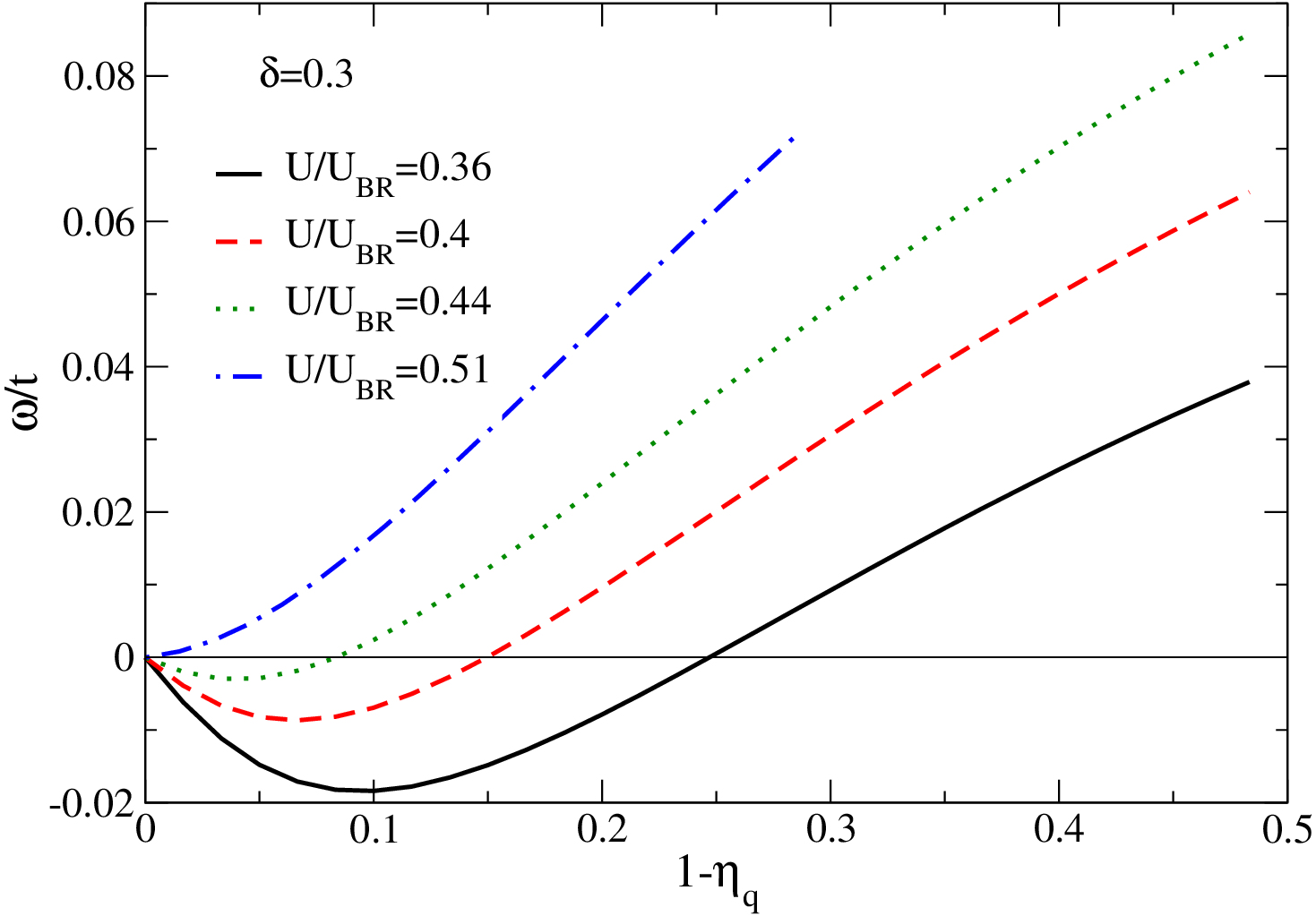}
\end{center}
\caption{(color online) Magnon dispersions for the (fully polarized) 
ferromagnet in the half-hypercubic lattice for $\delta=0.3$. At this doping
the FM solution is stable for $U>0.98 U_{BR}$.}
\label{fig35}
\end{figure}

We consider the hypercubic lattice and the  half-hypercubic
lattice. The latter  can be obtained from the hypercubic lattice
by removing all the even sites. For the hhc case  we consider hopping
restricted to  $t_{ij}=t/{\cal D}$ for   nearest neighbor sites and 
$t_{ij}=t/(2{\cal D})$ for 
next-nearest neighbor sites\cite{uhr96}. For the hypercubic lattice we
keep only  nearest-neighbor hopping  $t_{ij}=t/\sqrt{2\cal D}$. In the latter case it has been shown
\cite{mul89} that all the momentum dependence of correlation functions 
is contained in the factor
\begin{equation}
\eta_{\bf q} = \frac{1}{\cal D}\sum_{i=1}^{\cal D} \cos(q_i),
\end{equation}
which can be parametrized by a value between $1$ and  $-1$ corresponding
to a scan along the zone diagonal from ${\bf q}=(0,...,0)$ to 
${\bf q}=(\pi,...,\pi)$, respectively. On the other hand for the
half-hypercubic lattice the Brillouin zone is cut in half and the momentum
dependence is contained in the factor $\eta_{\bf q}^2$ which varies from 1
to 0 corresponding to a scan from ${\bf q}=(0,...,0)$ to 
${\bf q}=(\pi/2,...,\pi/2)$, respectively\cite{mul89}.

\begin{figure}[tbp]
\begin{center}
\includegraphics[width=7cm,clip=true]{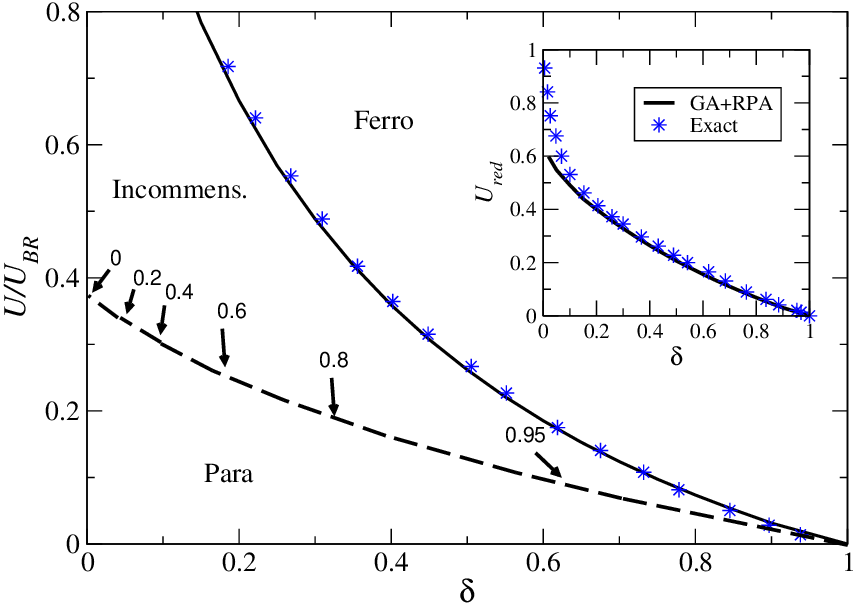}
\end{center}
\caption{(color online) Full line: limit of stability of the 
FM phase
  within the Gutzwiller wave function for the hhc lattice.  
The asterisks are the exact result from Ref.~\onlinecite{uhr96}. 
 In the inset a scaled representation of the  FM stability
  limit is shown with  $U_{red}=U/(U+U_{BR})$ and $U_{BR}=6.86t$.
Dashed line: limit of stability of the PM
 phase.  The arrows indicate the
  value of $\eta_{\bf q}$ parameterizing the momentum of the unstable
  mode. }\label{hhc}
\end{figure}

The hhc lattice is non-bipartite, therefore antiferromagnetism is frustrated.
In addition, it has a density of states with a square root divergence
at a lower band edge that makes ferromagnetism stable for large $U$
and large $\delta$. Here $\delta\equiv 1-n$ is the density
deviation from half-filling and  $n$ is the density. 
For convenience we define the Brinkmann-Rice critical $U$ value 
$U_{BR}\equiv-8\overline\epsilon$ with $\overline\epsilon$ the
noninteracting kinetic energy at half-filling \cite{vol84}.

Fig.~\ref{fig35} shows the transverse magnon dispersion computed on
top of a saturated FM solution at
$\delta=0.3$ in the hhc for different values of $U$ which as expected
is linear (quadratic) in $1-\eta_{\bf q}$ ({\bf q}) for small 
${\bf q}$ and shows a Goldstone mode at ${\bf q}=0$. As $U$ is decreased 
the excitation energy becomes negative first at ${\bf q}=0$ and then
also at finite momentum signaling an instability due to a change of
sign in the stiffness of the magnetization. 
Since this is a transverse excitation the ferromagnet tends 
to become a spiral.

In Fig.~\ref{hhc} we show the limit of stability of the saturated 
FM phase thus obtained (full line) and the exact computation by
Uhrig (asterisks). Only in the region where bound states appear
for small $\delta$ \cite{uhr96} significant deviations appear as shown
in the inset which indicates that the GWF energy  is quite accurate.

We also show in  Fig.~\ref{hhc} the stability limit of the
PM phase (dashed line). 
The first excited state of the paramagnet is a
triplet with z-component of the spin $m=-1,0,1$.
 Spin rotational invariance requires that the longitudinal  ($m=0$)
 and transverse ($m=-1,1$) excitations are degenerate and indeed all three 
become soft at the boundary  at some finite ${\bf q}$. In the figure we
indicate by arrows the value of $\eta_{\bf q}$ for the unstable modes. 
The instability occurs at $\eta_{\bf q}\rightarrow 0$ for small
$\delta$, consistent with the wave vector ${\bf q}=(\pi/2,...,\pi/2)$
 and gradually shifts to  
$\eta_{\bf q}=1$ for $\delta=1$ as expected for the instability
toward the FM phase (${\bf q}=0$).

Both instability lines indicate that there is an 
IC phase intermediate between the FM phase 
and the PM
phase. This does not show up in the 
DMFT numerical study of Ref. \cite{ulm98}
because it is restricted to ${\bf q}=0$
instabilities. The present FM boundary
practically coincides with the $T=0$ extrapolated results of 
Ref.~\cite{ulm98} while a substantial portion of the 
PM range of Ref.~\cite{ulm98}  is instead occupied 
by IC phases in the present computation.

\begin{figure}[tbp]
\begin{center}
\includegraphics[width=7cm,clip=true]{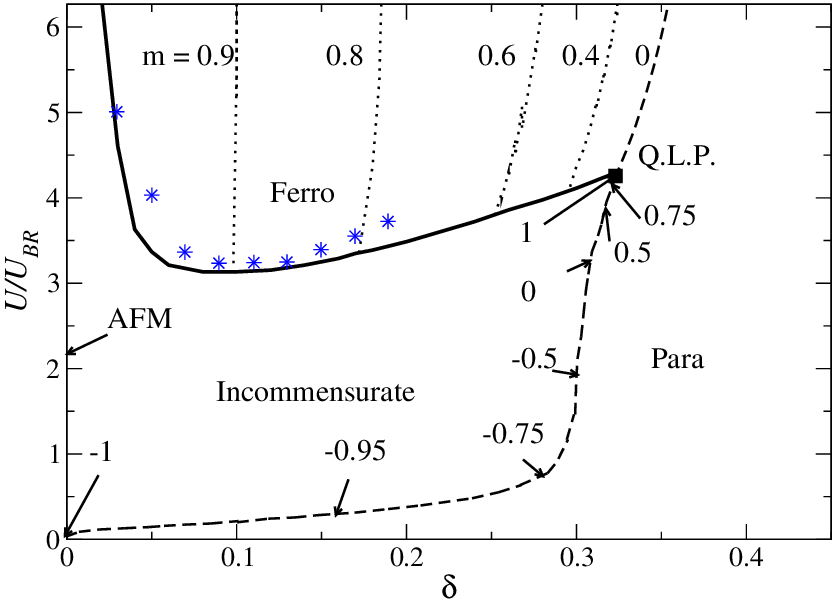}
\end{center}
\caption{(color online) Limit of stability of the FM (full line)
  and PM (dashed line) phases for the hc lattice.  
Arrows indicate the value of $\eta_{\bf q}$ for the unstable
  mode of the PM phase. The stars correspond to the limit of stability of the FM phase  
obtained in Ref.~\protect\onlinecite{obe97} within an approximate
solution of the DMFT equations. Dotted lines are levels of constant
  magnetization. Energies are in units of $U_{BR}=6.38t$.}\label{hc}
\end{figure}

Encouraged by these results which show the accuracy of the
Gutzwiller variational energy to determine the phase boundaries, 
we consider  the hc lattice. The limit of
stability of the FM phase and of the PM phase are
shown in Fig.~\ref{hc}.  
Since the hc lattice has perfect nesting the instability
towards antiferromagnetism ($\eta_{\bf q}=-1$) sets in at $U/t=0$ for
$\delta=0$ \cite{faz90,don95}. For $\delta>0$ an IC phase
 emerges  with $\eta_{\bf q}$ gradually increasing. 
  Decreasing $U$ at finite $\delta$ the FM phase 
becomes unstable at the full line
due to a change of sign in the stiffness of the magnetization as for the 
hhc lattice. Finally at the
  QLP the
 PM instability mode becomes uniform   
($\eta_{\bf q}=1$) and one has a transition to a 
weak (partially polarized) FM phase. 

The dotted lines 
in Fig.~\ref{hc} are level curves of constant magnetization $m$.
The density of states of the $\cal D=\infty$ 
hypercubic lattice is a Gaussian. Because of the Gaussian tails an
infinite splitting of the minority and majority bands,
  i.e. $U\rightarrow \infty$, is required to
produce a fully polarized state.  For finite $U$ one has 
a magnetization of the
FM state that increases from zero as $\delta$ decreases from the
PM boundary (dashed line) 
and becomes logarithmically saturated for small
$\delta$.

The limit of stability of the FM phase in Fig.~\ref{hc} 
is in good agreement with the
FM phase boundary for the hypercubic lattice
obtained in a DMFT study based on a large $U$ expansion and 
the no crossing approximation as impurity solver (asterisks)\cite{obe97}.
This again points to a good accuracy of the GWF.

The magnetic phase diagram of the  $\cal D=\infty$ 
hc lattice has been studied at finite temperature by Freericks and
Jarrell using DMFT \cite{fre95}. The finite temperature instabilities are 
towards a commensurate AFM and become IC only at the lowest
temperature which does not allow for a reliable $T=0$ extrapolation of
the PM-IC phase boundary. If we restrict to 
the AFM instability of the GWF we find an instability line in the 
$U-\delta$ plane (not shown) which is in good agreement with the $T=0$
extrapolated phase diagram of Ref.~\cite{fre95}. This is probably due
to the fact that the extrapolation is dominated by the high temperature
commensurate data. In the present phase diagram the IC phase
occupies a larger region. 

At the QLP the IC, PM and FM phases
meet. This behavior is the quantum analogue of the classical LP observed in 
 MnP where the same phases meet but at finite temperature and in the
 presence of an external field \cite{bec80}. 

On general grounds regarding the proximity of  ${\cal D}=3$ 
systems to infinite dimensional systems\cite{geo96} we expect the
phase diagram of Fig.~\ref{hc} to resemble that of the  
cubic lattice. It is then worth
speculating on the behavior of a QLP shall it occur at finite
${\cal D}$. In analogy with a classical LP also for a 
QLP we expect anomalous critical behavior.

The  $T=0$ PM-FM transition and PM-IC transition are the first
studied examples  
of quantum critical behavior \cite{mor73,her76,mil93,voj97,chub04,sachbook}. The
transition is characterized by a dynamical exponent  
$z=3$ ($z=2$) for the PM-FM (PM-IC)  
and the upper critical spatial dimension is $4-z$.  Therefore in many
cases of interest the theory is Gaussian.  
It was pointed out that in the case of
FM transitions because the magnetization is a conserved
quantity long-range interactions appear and the critical 
exponents change \cite{voj97}.

At the QLP the critical
behavior is expected to be different. Neglecting long-range
interactions the effective action close to the QLP can be written in
analogy to a classical LP and the FM QCP \cite{her76}: 
\begin{eqnarray}
  \label{eq:s}
  &&S=\frac12\int\frac{d^{\cal D}{\bf q}}{(2\pi)^{\cal D}} \nonumber \\
&&\frac1\beta \sum_n \left( r + c q^2 +
  D q^4 + \gamma \frac{|\omega_n|}q\right) {\vec \phi}({\bf q},\omega_n).{\vec\phi}(-{\bf q},-\omega_n)\nonumber\\
&&+u  \int d^{\cal D}{\bf x} \int_0^\beta 
d\tau [{\vec \phi}({\bf x},\tau).{\vec \phi}({\bf x},\tau)]^2.
\end{eqnarray} 
Here ${\vec\phi}({\bf x},\tau)$ is a  vector order parameter defined in
space and imaginary time, ${\vec \phi}({\bf q},\omega_n)$ is the
Fourier transform and $\omega_n$ is a bosonic Matsubara
frequency.  $D$, $u$ and $\gamma$ are positive constants. 
The parameters  $r$ and $c$  are linear functions of $\delta$ and
$U$ close to the quantum critical point.  
At mean-field level the QLP is defined by  
$c=0$, $r=0$. At the FM-PM boundary $r=0$ and $c>0$ whereas at 
the PM-IC phase boundary $c<0$ and $r=c^2/(4D)$ and the unstable mode has
momentum $q=\sqrt{-c/(2D)}$. The FM-IC boundary is characterized by
$c=0$, $r<0$ thus $c$ changes sign at the boundary mimicking the change
of sign in the stiffness found in RPA.  

As mentioned above, far from the QLP the   
transitions have a dynamical exponent $z=2,3$. At the QLP the
dynamical exponent
changes to $z=5$ and the upper critical dimension becomes $8-z=3$.
Therefore, if a QLP exists at low ${\cal D}$ the quartic term is not irrelevant 
and the theory becomes non-Gaussian or Gaussian with logarithmic corrections
in marked contrast with Hertz theory. 
A full analysis must take into account singular contributions to the 
coefficients and goes beyond our present scope.

One can also envisage an
AFM QLP at which AFM, PM and IC phases meet. In this case the damping
term becomes momentum independent implying $z=4$ with an upper
critical dimension ${\cal D}=4$. Cuprates  may 
 be not so far from this phenomenology since at low temperatures and 
as a function of $\delta$ they show a transition from an 
AFM state to a paramagnet/superconductor which in addition often shows
IC behavior \cite{tra95}. Even more relevant for our discussion may be 
 the low temperature behavior of overdoped 
cuprates where the IC states turns into a
paramagnet.  It has been argued\cite{chakra} that the system is also close to
a FM instability  which suggests the proximity to a IC-PM-FM
QLP. Another interesting system is  CeCu$_{6-x}$Au$_x$ for which an
 AFM QLP where only one direction becomes soft has been proposed \cite{ram99}.
In all these cases, however, disorder effects appear to be quite relevant.  
Another system where one may hope to find a QLP are ultracold 
atoms in optical lattices. In this fascinating realization of the
Hubbard model one can fine tune the parameters in the Hamiltonian
which may allow for a full exploration of the phase diagram \cite{dua03}.   

To conclude, our comparison with  DMFT studies shows that the  GWF, which
played and still plays a fundamental role in understanding strong
correlation, is a very convenient and accurate
tool to study the ${\cal D}\rightarrow \infty$ magnetic phase
diagram of the Hubbard model. We find that quite generically
incommensurate phases appear as intermediate phases between
strongly polarized FM phases and PM phases. This is interesting
because, as mentioned in the introduction, IC phases are observed in a
variety of systems in physical dimensions. 
For the hc lattice we find that the weakly polarized FM and the PM
 phase have a common phase boundary and both meet with the IC phase at
 a new QLP at which the magnetic order parameter is critical and the stiffness
simultaneously vanishes. We speculate that an analogous QLP may exist in
lower dimensional systems and we anticipate anomalous
quantum critical behavior.  

\acknowledgments We thank  C. Di Castro, C. Castellani,
S. Caprara, M. Grilli and M. Capone for useful discussions and G. Uhrig for
a critical reading of the manuscript. F.G. and G.S. acknowledge financial
support from the Deutsche Forschungsgemeinschaft.

\end{document}